\documentclass{aa}  
\usepackage{array}
\newcolumntype{P}[1]{>{\centering\arraybackslash}p{#1}}
\newcolumntype{M}[1]{>{\centering\arraybackslash}m{#1}}

\graphicspath{{./}{figures/}}
\usepackage{float}
\usepackage{graphicx}
\usepackage{txfonts}
\usepackage{caption}
\usepackage{subcaption}
\begin{document}

   \title{Improved Type III solar radio burst detection using congruent deep learning models}

   \author{J. Scully
          \inst{1}
          \and
          R. Flynn\inst{1}\and P. T. Gallagher\inst{2} \and E. P. Carley\inst{2} \and M. Daly\inst{1}
          }

   \institute{Department of Computer and Software Engineering,
            Technological University of the Shannon,
            Athlone, Ireland
              \email{j.scully@research.ait.ie}
         \and
             Astronomy \& Astrophysics Section, Dublin Institute for Advanced Studies, Dublin, Ireland.
             }

   \date{March 14, 2023}

 
  \abstract
   {Solar flares are energetic events in the solar atmosphere that are often linked with solar radio bursts (SRBs). SRBs are observed at metric to decametric wavelengths and are classified into five spectral classes (Type I--V) based on their signature in dynamic spectra. The automatic detection and classification of SRBs is a challenge due to their heterogeneous form. Near-realtime detection and classification of SRBs has become a necessity in recent years due to large data rates generated by advanced radio telescopes such as the LOw Frequency ARray (LOFAR). In this study, we implement congruent deep learning models to automatically detect and classify Type III SRBs. We generated simulated Type III SRBs, which were comparable to Type IIIs seen in real observations, using a deep learning method known as Generative Adversarial Network (GAN). This simulated data was combined with observations from LOFAR to produce a training set that was used to train an object detection model known as YOLOv2 (You Only Look Once). Using this congruent deep learning model system, we can accurately detect Type III SRBs at a mean Average Precision (mAP) value of 77.71\%.}

   \keywords{solar radio bursts, deep learning, generative adversarial network, YOLO}

   \maketitle
%

\section{Introduction}
Solar flares are the most intense explosive events in the solar system \citep{Lin_2011}. The accelerated particles emit light across the electromagnetic spectrum, from gamma rays to radio waves. High-intensity radio emission characterizes solar radio bursts (SRBs), which manifest as complex signals in dynamic spectra. Based on the structure of their dynamic spectra, SRBs are divided into five categories, ranging from Type I to Type V \citep{Pick}. Because Type III bursts can occur hundreds of times each day, detecting them and understanding their spectral features is a computational challenge. This problem has become even more difficult in recent years, with the development of technologies such as the LOw-Frequency ARray \citep[LOFAR;][]{Van_Haarlem}, which provides high-volume data streams (up to 3 Gb/s at a single station) of radio burst observations that need to be classified accurately in real-time. With the development of LOFAR for Space Weather \citep[LOFAR4SW;][]{Carley}, a system update aimed at autonomously monitoring solar radioactivity, the need for automated data pipelines for solar radio bursts has become more immediate. Presently, software pipelines for autonomously detecting SRBs will be an essential component of such a system. The research presented in this paper identifies deep learning as an important component of such a pipeline.

Recent work on object detection algorithms, such as You Only Look Once \citep[YOLO;][]{YOLO}, classification algorithms, such as Support Vector Machines \citep[SVM;][]{SVM} and Random Forest \citep[RF;][]{RF}, has shown the need for a high-quality simulated training set to improve the accuracy and robustness of algorithms when classifying and detecting Type III SRBs \citep{Carl}. The work presented in this paper shows the crucial role that Generative Adversarial Networks (GANs) \citep{Gans} play in generating simulated images for such training sets.

Creating training sets for SRB classification using conventional techniques frequently involves the tedious effort of combing through large data archives to locate and collate pertinent Type III SRB images. Generating simulated data significantly reduces this task through generation of data that not only look like SRBs but also creates it in volume and in a short period of time. Previously, a significant number of SRB-like images were produced using parametric modeling \citep{parametric}. This method produced Type III SRBs that were random in number, grouping, intensity, drift rate, heterogeneity, start-end frequency, and start-end duration, all of which are traits of a Type III. However, compared to Type IIIs observed in daily observations, these images lacked realism.

Several non-machine learning attempts have been made to automatically classify SRBs in dynamic spectra. Current algorithms implement the Hough or Radon transform methods as a way of recognizing specified parametric shapes in images \citep{Lobzin}. Depending on the type of radio burst being categorized, these algorithms can reach up to 84\% accuracy. Other methods include Constant-False-Alarm-Rate detection \citep{CRAF}, which is essentially the detection of radio bursts in dynamic spectra employing de-noising and adaptive threshold. The method works well with various types of radio bursts and has been reported to have a 70\% accuracy.

Through the application of multi-modal deep learning to a spectrogram at millimetric wavelengths, deep neural networks have been shown to be highly successful at detecting SRBs \citep{Multimodal}. The technique combines auto-encoders and regularization to achieve an accuracy of 82\% in burst detection, but it has not been applied to metric wavelengths (the LOFAR range), where bursts can have much more intricate geometries.

Generative deep learning models, such as Deep Convolutional Generative Adversarial Networks (DCGANS) \citep{DCGAN_SRB}, are playing a crucial role when classifying SRBs. The system is modified to convert a GANs generative type network into a classification technique. This method successfully classifies Type III solar radio bursts with an accuracy of between 89-92\% when used with LOFAR metric wavelengths. 

To identify SRBs, researchers have recently started using object detection algorithms such as Faster R-CNN \citep{Faster_RCNN_SRB}. With an average precision (AP) of 91\%, this deep learning neural network was demonstrated to be accurate at extracting minor aspects of SRBs. However, the model lacks the capability of real-time detection, demonstrating a maximum of 17 frames per second on open-sourced datasets such as COCO \citep{Faster_RCNN}.

Fast Radio Bursts (FRBs) and Search for Extra-Terrestrial Intelligence \citep[SETI;][]{Seti} is another area where machine learning is being used in conjunction with radio interferometer observations. SETI employs machine learning to look for FRBs in planetary systems using its Allen telescope array. SETI uses a deep Convolutional Neural Network \citep{CNN}, known as ResNet \citep{He_Zhang}, to extract irregular high-frequency FRB spikes from within the dynamic spectrum, highlighting regular noise frequencies and Radio Frequency Interference (RFI). This model produced a recall score of 95\%.

Conor \& van Leeuwen use deep neural networks to extract and classify features of FRBs within observations \citep{Connor}. The research uses both simulated and real single Galactic pulsars to obtain a dataset for training the CNN. In this instance, due to the scarcity of FRBs, the training set was dominated by simulated bursts. They simulated most of their true positives but used only false positives that were generated in real surveys and labelled by eye. CNNs have been applied to this scarce data with above 99\% accuracy when classifying such phenomena. This FRB extraction method illustrates how robust a CNN can be when using a hybrid dataset containing both real and simulated data.

With recent research moving to deep CNNs and object detection for classifying and detecting radio frequency phenomena, we decided to further investigate these topics. For object detection, CNNs come in a variety of different flavours, including YOLO, Single Shot Detectors  \citep{SSD}, Region-CNN (R-CNN) \citep{R-CNN}, Fast R-CNN  \citep{Fast_RCNN}, and Faster R-CNN \citep{Faster_RCNN}. YOLO has been the only algorithm to deliver high accuracy and real-time detections on datasets, although the other approaches listed have shown to be quite successful for object detection, just not in real-time.

In our previous research, we adapted the YOLO algorithm to detect Type III SRBs \citep{Scully_Flynn_Carley_Gallagher_Daly_2021}. Using this configuration we obtained an accuracy score of 82.63\%. However, one score we could not obtain then was mean Average Precision (mAP), which determines how precisely the algorithm can locate a certain object in an image, in this case, Type III SRBs. We noted some key areas that we could improve on to allow us to measure mAP, in particular the simulated training set.

In this paper, we apply multiple deep-learning methods to the problem of SRB simulation, detection, and classification. We use GANs to create a simulated training set of images comparable to real observations, which is then fed into YOLO to precisely detect and classify Type III SRBs. 

The paper is organised as follows, In Section 2, we describe how our data is gathered with LOFAR and the phenomenon that is an SRB. In Section 3, we introduce the deep learning technique of convolutional neural networks and how we manipulate this deep learning technique for SRB simulation, detection, and classification, and the significance of SRB simulation techniques. In Section 4, we introduce GANs, and how they can improve SRB simulation. In Section 5, we introduce YOLO, our second deep learning method for SRB detection and classification, its architecture, and the datasets on which YOLO is trained and evaluated. In Section 6, we visualise and discuss the results produced by YOLO and how the introduction of GANs for simulating SRBs can generate a mAP score.

\section{LOFAR and Solar Radio Bursts} \label{sec:LOFAR}
\subsection{LOw Frequency ARray}
\begin{figure}[ht]
    \centering
    \includegraphics[width=1\linewidth]{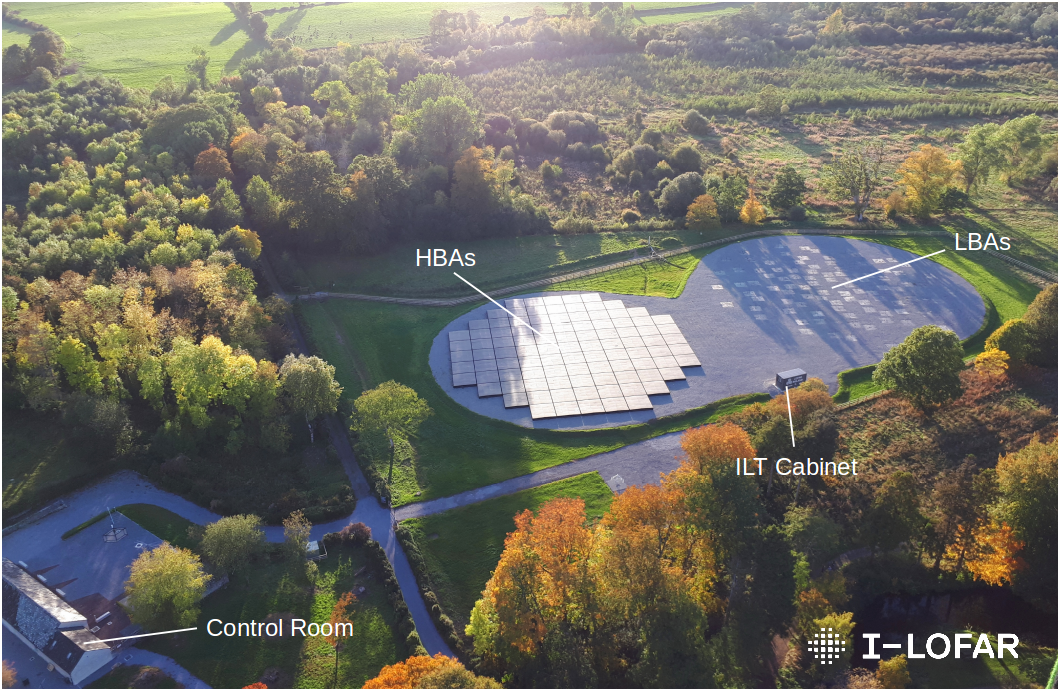}
    \caption{The Irish Low-Frequency Array station IE613 (I-LOFAR) at Birr Castle, County Offaly. Coaxial cables are used to transport data from the Low Band Antennas (LBAs) and High Band Antennas (HBAs) to the ILT Cabinet (center right), where it is amplified, filtered, and digitalised. Data is sent in international mode at a speed of $\sim$3.2~Gbps to Groningen, Netherlands. In the I-LOFAR Control Room, data is processed using REALTA in local mode (bottom left)\citep{Murphy}.}
    \label{fig:ILOFAR}
\end{figure}
The data for this study came from the LOFAR radio interferometer, which was erected in the north of the Netherlands and across Europe, and includes Ireland's I-LOFAR station, which is shown in Figure 1. For monitoring the radio universe in the comparatively uncharted low-frequency range of 10-240 MHz, LOFAR offers a special assortment of observational modes. Several radio astronomical objects can be observed simultaneously by LOFAR, which can also operate several stations at once. The system can operate as a multi-station, very long baseline interferometer or each station can function independently as a telescope.

LOFAR antenna stations provide the same basic functions as standard interferometric radio dishes. These stations feature a large collecting area and high sensitivity, as well as pointing and tracking capabilities, similar to typical radio dishes. Unlike traditional radio dishes, LOFAR stations physically don't move, instead, the system combines signals from individual antennae to construct a phased array using a combination of analog and digital beam-forming techniques, making it more flexible and agile. Rapid telescope re-pointing and multiple, simultaneous observations from a single station are made possible by station-level beamforming. Then, a central processing unit can receive the stations' digitized, beam-formed data and correlate it for imaging and observation analysis.

Single-station beamformed data is typically compiled into a dynamic spectrum with a time resolution of 5 microseconds and a frequency resolution of 195 kHz. The 488 frequency channels that make up the dynamic spectra allow for the recording of data at a rate of several Terabytes (TB) per hour. The REAL-time Transient Acquisition backend (REALTA), a high-performance computing system for processing and storing unprocessed beamformed data has recently been developed by the I-LOFAR team, \citep{Murphy}. Due to the high volume of data, automated algorithms are required to sort and classify any phenomena of interest. In our case, the classification and detection of SRBs is the primary goal. In this study, we constructed training sets on which GANs and YOLO could be trained and assessed against observations from I-LOFAR. 

\subsection{Solar Radio Bursts}
SRBs are usually investigated in dynamic spectra and are divided into five major spectral classes, spanning from Type I to Type V, based on their shape, frequency, and time length. The intricate characterisation of these radio bursts, however, makes classification extremely challenging. When employing machine learning approaches to categorize such occurrences, the data that the classification algorithms are trained on is crucial. SRBs are regularly observed in dynamic frequency versus time spectra.

The most frequent SRBs are Type IIIs, which are short-lived and appear in dynamic spectra as a vertical bright strip, as illustrated in Figure 2. The numerous diverse forms a Type III might assume within this vertical strip makes the task more complicated. They can appear as being smooth or patchy, faint, or intense, superimposed on other radio bursts, freestanding or in groups, or immersed in strong RFI. For this research, we used Type IIIs in the frequency range of 10-100 MHz, where they generally occur as a vertical stripe.
\begin{figure}[ht]
    \centering
    \includegraphics[width=\columnwidth]{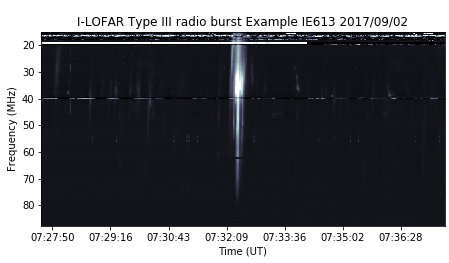}
    \caption{Example of a dynamic spectrum showing a real Type III solar radio burst between 20-90MHz. Notice the Type III's vertical strip-like shape and short duration in time, lasting only a couple of seconds.}
    \label{fig:giant_pulse}
\end{figure}
\section{Deep learning and data simulation} \label{sec:style}
\subsection{Convolutional neural networks}
 A CNN is a deep-learning neural network that is specifically designed for visual feature recognition. It is distinguished from a standard NN by the presence of numerous convolutional layers. These layers effectively do image filtering to produce image intensity gradients. Each filter produces a different gradient and is responsive to particular shapes. For example, the first layers of the CNN may contain filters that respond to horizontal or vertical lines, curves, and other simple geometric shapes. After the first filters are applied, the CNN then applies a max-pooling layer, which is a type of summing and downsampling. Another convolutional layer, max-pooling, and so forth follow. Max-pooling works by allowing deeper layers of the network to access larger and larger portions of the image. While simple geometries may be responsive to the early layers, with max-pooling, the subsequent layers become responsive to more complex shapes made from these geometries, such as circles, triangles, and complex polygons. The max-pooling and convolving can continue until the deepest levels of the network react to recognizable aspects, for example, facial characteristics. 
 
 In our research, the network's final layers respond to SRBs and their precise structure. Depending on the complexity of the network, the number of convolutional and max pooling layers vary, but the final result is a single image vector representation. This vector is then used to classify the image. Similar to smaller neural networks, a CNN's weights and biases (including those of the convolutional layers) must be trained. Given the size of the networks, fitting tens of thousands or even millions of parameters may be necessary for the most complex networks. To avoid under- or overfitting \citep{overfit}, the number of training instances should at least be in or close to the same order of magnitude as the number of weights and biases in the network. A resulting problem is the lack of accessible databases with tens of thousands of images to train phenomena like radio bursts. There are two ways to address this problem: (i) simulating a large number of training samples for the network, or (ii) using a method called transfer learning \citep{transfer}, in which we take an advanced and powerful CNN that has already been trained on millions of images of generic scenes (containing, among other things, everyday objects like cars and people) and retrain a smaller portion of the network on our specific set of data (radio bursts). Transfer learning is based on the notion that the general forms learned by the CNN from ordinary items may be recycled for new objects, with studies showing that this is possible even for images that are wholly morphologically unlike, such as cars and solar radio bursts \citep{transfer}. 
 
In this research, we implemented congruent CNN architectures to address the data problem. In deep learning, congruence refers to the similarity or consistency between different parts of a model or different models. It also refers to the similarity between the weights of different models trained on the same task, or the consistency of the outputs of different models when presented with the same input. In the case of GANs, we created two CNN architectures, with one being inverted and pitched them against each other to generate real simulated Type III SRBs. We then used transfer learning along with the simulated dataset to train an elongated CNN architecture YOLO to detect Type III SRBs.
 
\subsection{SRB simulation}
Simulating SRBs has been essential for generating large datasets for training detection and classification algorithms. It essentially eliminates the time-consuming task of searching large data archives to find good, clean images suitable for a training set. Simulating data also gets rid of the need for cleaning archival data of any artefacts such as RFI. We originally used parametric modelling to simulate SRBs in previous studies, in which polynomials were used to create overall Type III shape in dynamic spectra, with skewed Gaussians used for their time dependant intensity profile at each frequency. Using this method we created Type III radio bursts with a random number, grouping, intensity, drift rate, heterogeneity, start-end frequency, and start-end time. We placed the bursts in front of a background of synthetic and random RFI channels, as shown in Figure 3. 
\begin{figure}[ht]
    \begin{subfigure}[b]{0.24\textwidth}
         \centering
         \includegraphics[width=\textwidth]{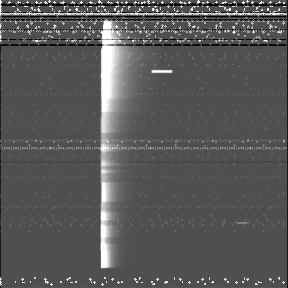}
         \caption{Parametric Type II}
         
     \end{subfigure}
     \hfill
     \begin{subfigure}[b]{0.24\textwidth}
         \centering
         \includegraphics[width=\textwidth,height=4.43cm]{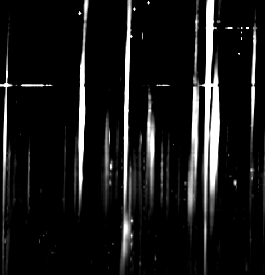}
         \caption{Real observed Type II}
         
     \end{subfigure}
    \caption{Parametric modelling simulation (a) compared to a real Type III observation (b). The parametric modelling method fails to simulate activity seen in a real observation such as small, faint Type III bursts or interference like embedded RFI.}
\end{figure}

When training YOLO, this strategy has a number of advantages, including the ability to generate enormous amounts of automatically labelled Type III data in a short amount of time. Using parametric modelling, we were able to construct a dataset of 80,000 images of Type III SRBs for our previous YOLO model. However, this training set created many issues when it came to testing the model's robustness. The first issue is that this data must be accurately labelled for training; YOLO needs to see what it's looking at because it's a supervised learning technique that requires a labelled dataset in order to identify a class, in this instance Type III SRBs. The automatic labelling system in the parametric modelling saved a lot of time, but there was no change in the Y-axis or height variables of these labelled bounding boxes, in other words, the height variable became overly saturated with the same static values in the training set, therefore the final YOLO detections had the same height variation as the training set, as seen in Figure 4. As a result, we were unable to calculate the localized precision of our model.
\begin{figure}[ht]
    \centering
    \includegraphics[width=\columnwidth,height=6.5cm]{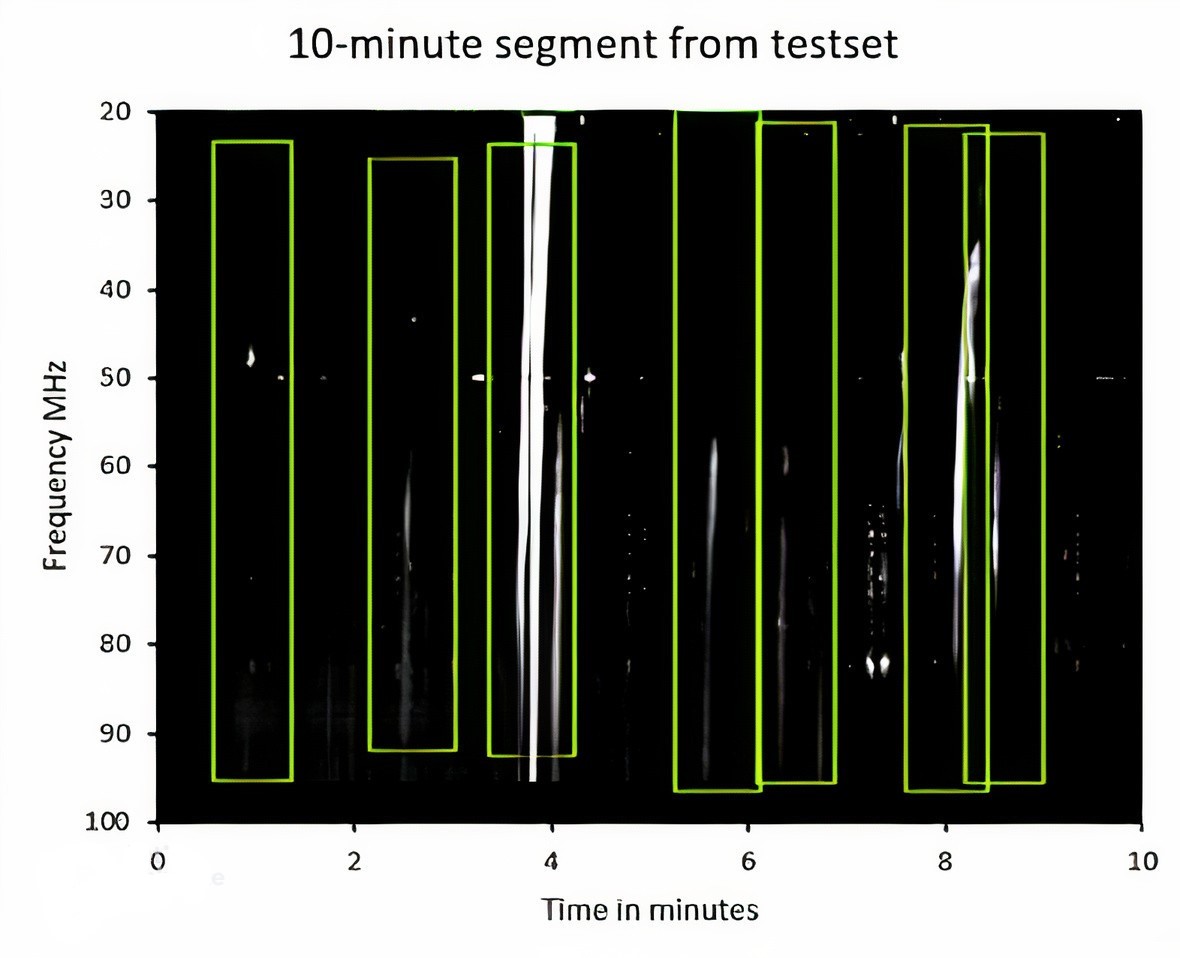}
    \caption{A 10-minute segment from the testset of our previous model attempt. The lack of variation and over-saturation in the Y-axis or height variable in the training set meant that when our previous YOLO model was evaluated, the bounding box predictions (highlighted in green) in the test set had no variation in the Y-axis or height variable no matter the size of the Type III.}
    \label{fig:prev_YOLO}
\end{figure}

The second issue we experienced with the parametric modelling approach was the model's lack of realism. While it provided many possibilities in terms of position, grouping, and overall shape and intensity, it did not provide us with the exact shape and intensity variation that we would see in a real observation. 

With the introduction of GAN, we were able to generate Type III SRB simulations that were realistic. This allowed us to create SRBs that were similar to those seen in actual I-LOFAR observations, thus removing the need to trawl through data archives for the appropriate images for the training set. They also offered the Y-axis variation that YOLO needed to make localized Type III SRB detections.

\section{Generative Adversarial networks} \label{sec:GANs}
\begin{figure*}[ht]
    \centering
    \includegraphics[width=1\linewidth,height=9cm]{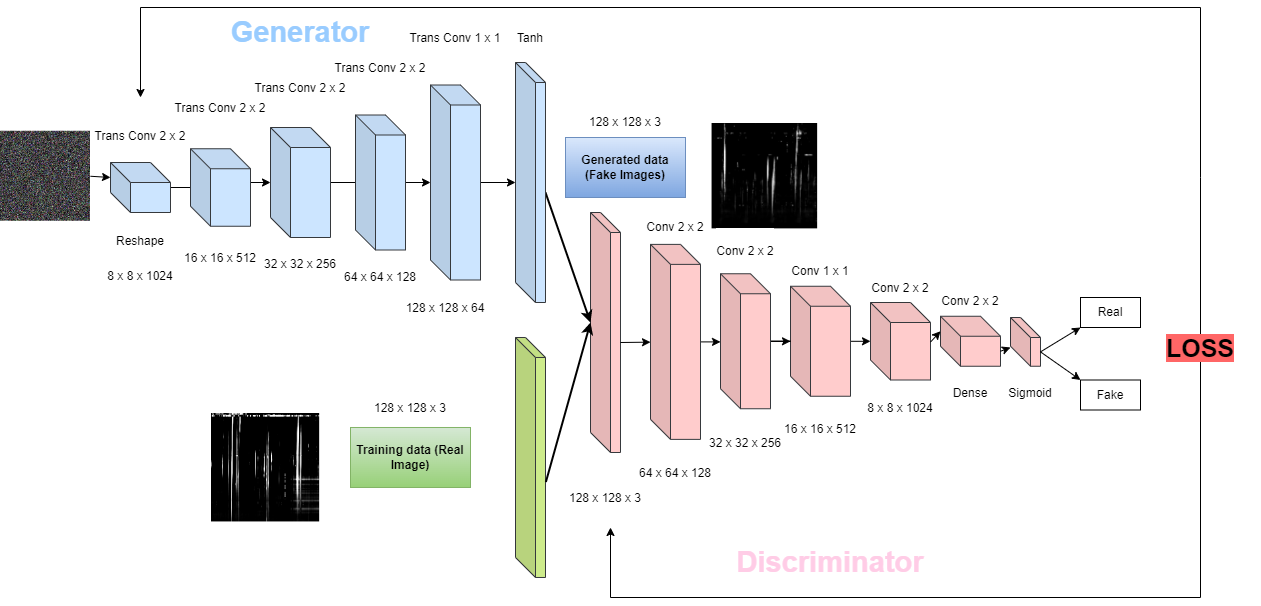}
    \caption{An illustration of a GAN architecture for producing Type III SRB data in simulation.
The purpose of the generator is to accept random input values (noise) and generate an image from them using a deconvolutional neural network. To upsample the data, we employ 5 transpose convolutional layers with a variation of 2 x 2 and 1 x 1 strides. In order to accommodate for small negative values when the input is less than zero, we use Batch normalization and Leaky ReLu activation after each transposition layer. The Tanh activation function is used at the top layer because, while creating images, they are often normalized to fall between [0,1] and [-1,1]. The discriminator is then fed a batch of real training data or fake training data, depending on the training stage, then downsampled using 5 convolutional layers with a combination of 2 x 2 and 1 x 1 strides. Each convolutional layer is followed by the use of Batch normalization once more, followed by ReLu activation, which changes all negative values to 0. The Sigmoid activation function, used at the top layer, normalizes the output in the [0, 1] range.}
   
\end{figure*}
GANs are a type of generative modelling that makes use of CNNs. Generative modelling is an unsupervised learning task that automatically finds and learns regularities or patterns in input data. The model created may be used to produce or output brand new instances that have similiar attributes or features from the original dataset input. The generator model is an inverted CNN, used to generate new simulated instances, and the discriminator model, a binary classifier that attempts to classify examples as real or fake (generated), are the two sub-models that are trained as part of the GANs framework, see Figure 5. These two sub-models are trained in an adversarial zero-sum game until the discriminator model is tricked approximately half of the time, indicating that the generator model is producing believable examples. To put it simply, GANs let us generate incredibly realistic new data that is based on pre-existing data.

\subsection{Type III Generation}
The GAN was trained to create simulated Type III SRBs. The training set consisted of 2,763 real Type III images that I-LOFAR obtained by merging several observation days, with each observation day broken up into 10-minute chunks. The vertical strip shape of a Type III is visible in these 10-minute chunks. In terms of solar activity, these observation days alternate between active and relatively quiet. This data is then cleaned to generate images that are free of interference, such as embedded RFI. We don't need to label any data because the GAN is an unsupervised algorithm, so the images are fed directly into the algorithm for training. The GAN algorithm is quite computationally demanding as we are trying to produce images from just general noise or random vectors. Therefore, the system used to train the GAN included  two Nvidia Geforce RTX 2080 Ti GPUs connected via SLI, running Ubuntu 20.4.2 LTS on an AMD Ryzen Threadripper 1950x with 32GB of RAM. For the training configuration, we were using 90\% of GPU capacity for a variety of different epochs at a batch size of 32, see Figure 6.
\begin{figure*}[ht]
\centering
   \begin{subfigure}[b]{0.49\textwidth}
         \centering
         \includegraphics[width=\textwidth]{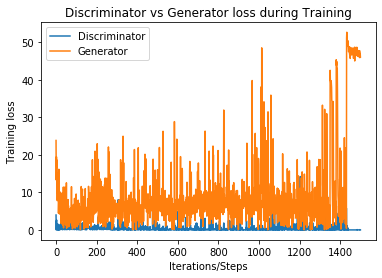}
         \caption{}
     \end{subfigure}
     \hfill
     \begin{subfigure}[b]{0.49\textwidth}
         \centering
         \includegraphics[width=\textwidth]{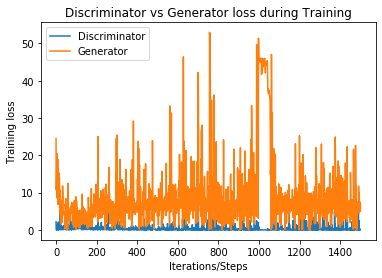}
        \caption{}
          \end{subfigure}
          \hfill
     \begin{subfigure}[b]{0.49\textwidth}
         \centering
         \includegraphics[width=\textwidth]{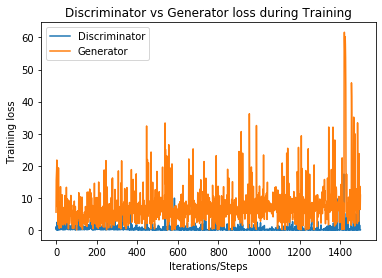}
    \caption{}
          \end{subfigure}
          \hfill
     \begin{subfigure}[b]{0.49\textwidth}
         \centering
         \includegraphics[width=\textwidth]{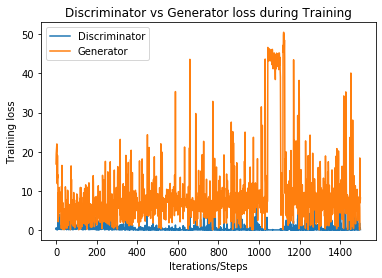}
         \caption{}
     \end{subfigure}
          
\caption{The loss error battle between the discriminator and the generator when generating Type IIIs. This illustrates the GAN's learning pattern. Notice how no instance of training is the same. One key feature when training GANs is convergence failure seen in plots (b) and (d) (when generator loss spikes) \citep{Goodfellow_2016}. This occurs when there is an inability to find the equilibrium between generator loss and discriminator loss. Images generated during this period are very poor and noisy.
\label{fig:Gans_training}}
\end{figure*}

During training, we produced 8 generated images after each epoch to create a collection of fake images. The GAN was trained numerous times, which allowed us to build a dataset of over 4,500 simulated Type III SRBs that were random in number, grouping, intensity, drift rate, heterogeneity, start-end frequency, and start-end time. The generated SRBs compared very well with real Type IIIs observed by I-LOFAR. We then filtered out noisy generated images produced when the generator error significantly spiked during training. The generated images were small at 128 x 128 pixels, so they were bulk-rescaled up to 256 x 256. To evaluate the images produced by the GAN, we use human perception as the most efficient way of evaluating these GAN-produced Type III SRBs \citep{Borji_2019}. We compared the GAN results along with parametric methods to real Type III SRBs observed by I-LOFAR in Figure 7. This generated data was then used as a hybrid training set for YOLO.
\begin{figure*}[ht]
\centering
    \begin{subfigure}[b]{0.30\textwidth}
         \centering
         \includegraphics[width=\textwidth]{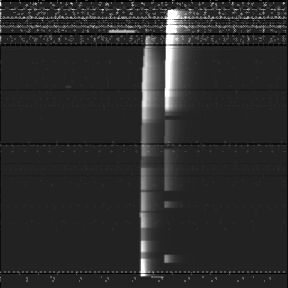}
         \caption{Parametric modelling}
     \end{subfigure}
     \hfill
     \begin{subfigure}[b]{0.30\textwidth}
         \centering
         \includegraphics[width=\textwidth,height=5.53cm]{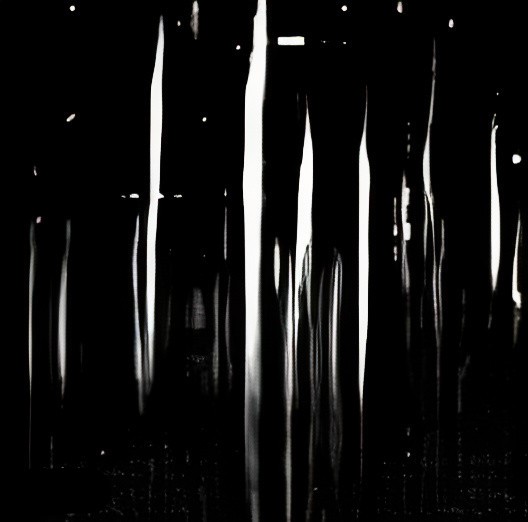}
        \caption{GANs}
          \end{subfigure}
          \hfill
     \begin{subfigure}[b]{0.30\textwidth}
         \centering
         \includegraphics[width=\textwidth]{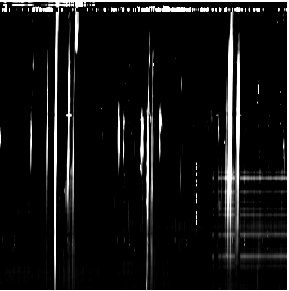}
    \caption{Real observation}
          \end{subfigure}
\caption{Parametric modelling and GANs compared to real Type III SRBs observed by I-LOFAR. GANs produce more realistic examples of Type IIIs compared to the parametric modelling method. These GAN-produced Type IIIs were combined with real Type IIIs observed by I-LOFAR to create a training set for YOLO.
\label{fig:GANs}}
\end{figure*}
\section{You only look once (YOLO)}\label{sec:YOLO}
Identifying what entities are present in a given image, and where they are located, is a computer vision problem known as object detection. Detecting objects in an image is more difficult than classifying them, as classifying only distinguishes between objects but not their exact positions in an image. Additionally, classification fails when applied to images with numerous objects. YOLO employs a different strategy. YOLO is a CNN that, depending on its configuration, does real-time object detection. The method divides the image into grid regions and predicts bounding boxes and probabilities for each grid zone using a single CNN (see Figure 8). Projected probabilities are used to weight these bounding boxes. Due to its high accuracy and real-time functionality, YOLO is well-liked.
The approach "only looks once" at the image since it only needs one forward propagation run through the neural network to provide predictions. Using YOLO, a single CNN can predict a variety of bounding boxes and class probabilities for those boxes. By using complete images for training, YOLO enhances detection performance. Because Type III SRBs are often short-lived ($\sim$0.1-3 seconds) and have a drift rate of 500 MHz$^{-1}$ in dynamic spectra \citep{typeoccurence}, we chose YOLO for this investigation. The fundamental benefit of YOLO is that it is very quick and can deliver accuracy that is virtually equivalent to Faster R-CNN \citep{Faster_RCNN}.
\subsection{Dataset}
The key feature of our updated YOLOv2 model from our previous work is the dataset. Instead of using the parametric modelling generated data, we used a hybrid dataset, which offered more realistic data for YOLO to train on. The hybrid dataset consisted of data generated by GANs and real observed data from I-LOFAR in a 50:50 split. This improved dataset of 6,732 images, with just over 60,000 Type III examples, is considerably smaller than the parametric modelling (80,000 images) training set. However, the improved dataset is more realistic and offers more robustness when testing on real I-LOFAR observations. There’s also less memory taken by the improved training set so it can be easily transferred without the risk of data corruption. The only constraint that comes with this approach is that this dataset needs to be manually labelled. Once the dataset was labelled, we had a training set consisting of 6,732 images for training YOLOv2.
\subsection{Model configuration}
After we created the dataset, we set up the model. To create the model in YOLOv2, we used a framework called Darkflow \citep{darkflow}, which is a TensorFlow python version of Darknet. As seen in Figure 8, YOLOv2 has 19 convolutional layers and 5 maxpool layers.
\begin{equation}
    filters = bounding (classes + coords)
\end{equation}
To optimize the model, the number of filters in the final convolutional layer was reduced (see function 1), and the sizes of the bounding boxes were adjusted using anchor values or bounding box dimensions. In our previous research, we originally set the height of the bounding box to a static 10-90MHz; this was due to the lack of Y-axis value variation seen in the parametric modelling approach. With our new and improved training set, we adjusted the anchor value ranges to 10-30MHz, 10-40Mhz, 10-50MHz, 10-60MHz, and 10-80MHz. This allowed us to capture most Type III sizes in terms of width and height detected by YOLOv2. The input size was also changed from 416x416 to 288x288 to achieve real-time frame rates. With this input size, the model can detect Type IIIs in real-time (90 frames per second) as it only detects one class in greyscale format \citep{YOLO9000}. 
\begin{figure*}[ht]
    \centering
    \includegraphics[width=1\linewidth]{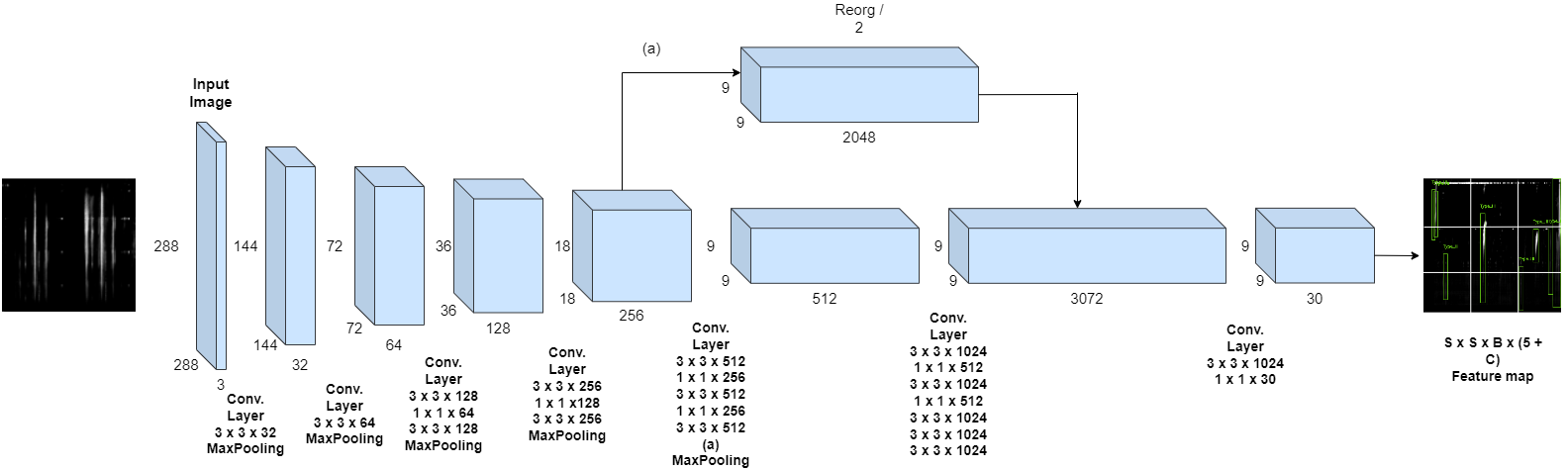}
    \caption{The Darknet-19 CNN architecture of YOLOv2. The Darknet-19 architecture consists of 19 convolutional layers and 5 maxpool layers. The Reorg layer combines both high- and mid-level features for better detection accuracy. In order to increase accuracy in YOLOv2, the fully connected layers of the CNN are eliminated, and K-means classification is used for detection and classification \citep{YOLO9000}.}
    \label{fig:GAN_exp}
\end{figure*}
\subsection{Training and Validation}
The YOLOv2 model was trained to detect and classify Type III SRBs. The training set employed a collection of 3,000 simulated GAN images and 3,763 real images. The data set included Type III samples with random start-end frequency, start-end time, drift rate, intensity, grouping, and inhomogeneity. A subset of the training set, the validation set, contained 1,500 Type III images produced by the GAN. The training and validation sets were both manually labelled which, although tedious, provides YOLO with precise instructions on what to train within the specified image. To fulfill the requirements of the Darkflows training set, these manually labeled training set images were converted into XML instructions. To determine whether the model was overfitting or underfitting, we had to construct the Darkflow framework to validate the training. Leaky ReLU was used as the activation function during training, and the model was trained using a learning rate based on the learning pattern of the model. If the model learns too quickly, the learning rate was updated, resulting in a smoother learning pattern. Stochastic Gradient Descent (SGD) was used to continuously update the learning parameters until convergence. We trained the YOLO model for 1,000 epochs at a batch size of 16. With this configuration, training took 7 days with both training and validation loss decreasing with every iteration, as shown in Figure 9.
\begin{figure}[ht]
    \centering
    \includegraphics[width=\columnwidth,height=5cm]{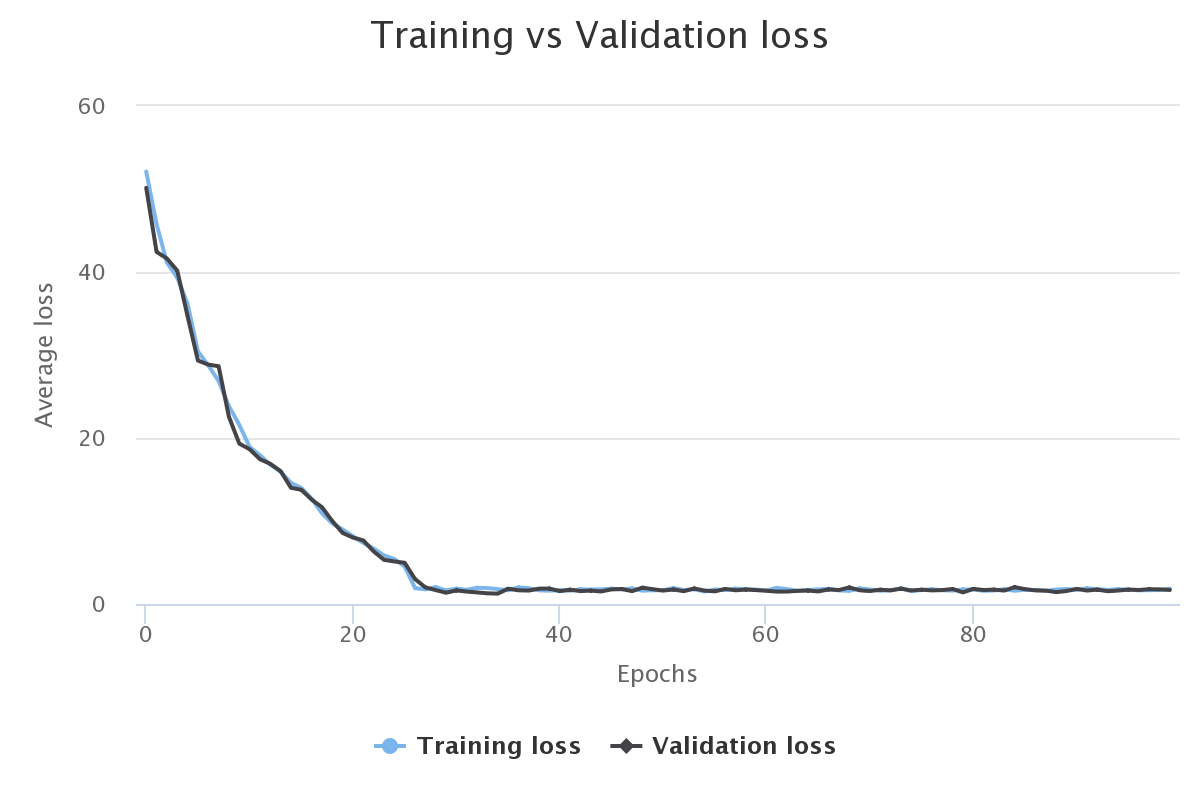}
    \caption{Comparing Training loss with validation loss illustrates how good YOLO is learning on the training set. It is also used to prevent the algorithm from overfitting. Each epoch represents 422 iterations, or when the dataset is passed forward and backward through YOLO once.}
    \label{fig:YOLO_learn}
\end{figure}
\subsection{Test set}
When testing our previous model, we chose a specific observation date, the 10th of September 2017. Although it provided it us with good benchmark results it never tested the robustness of our model as the observation was relatively uneventful. To test the model's robustness, a test set is needed that has a variety of Type III examples. Therefore, multiple observations from different dates in the I-LOFAR archive were mined to build a test set that contained examples of busy and quiet periods of solar activity. In order for colour not to be an influencing factor in the model's predictions, the image was converted to greyscale \citep{colour}. We concentrated our observations in the 10-90 MHz band as this is where the Type III's vertical strip shape can be seen. The observations were then divided into 10-minute intervals to provide a test set of 2,763 images that contained around 35,000 Type III solar radio bursts. We then precisely annotated our ground truth bounding box values. When a Type III was labeled, the appropriate bounding box coordinates were saved in an XML file for mAP for comparison of the ground truth coordinates with the predicted coordinates of the models. 
\section{Results}
The YOLOv2 model’s performance was measured using the test set described in Section 5.4. In our previous research, when evaluating the model’s performance, we viewed the model as a detection-classification problem. This was done by first predicting using the test set and then sifting through it searching for correctly identified Type III SRBs. Once a bounding box encompassed a Type III, we categorised it as correctly identified and then annotated a bounding box around the predicted bounding box. This meant we could only obtain a unit in which we represented our previous model’s performance results known as the f1-score. The f1-score calculates the balance between precision and recall, where precision refers to how accurately the model predicted an object's position and recall is the proportion of true positives to all actual objects. Then: \begin{equation}
    Precision = \frac{TP} {TP + FP}
\end{equation} 
\begin{equation}
    Recall = \frac{TP} {TP + FN}
\end{equation}
\begin{equation}
f1-score =  2 * \frac{Precision * Recall}{Precision + Recall}
\end{equation}
Although we could determine how accurate our model was, we could never calculate mAP (mean Average Precision), which is a metric used in computer vision for evaluating the performance of object detection and image classification algorithms. It calculates the average precision across all classes and is expressed as a fraction between 0 and 1. It takes into account both the number of true positive (TP) detections and false positive (FP) detections. A higher mAP score indicates a better performance of the algorithm. The TP and FP variables in mAP are determined by comparing the ground truth bounding box to the model's predicted bounding box, also known as Intersection over Union (IoU).

\begin{equation}
    Intersection\,over\,Union = \frac{Area\,of\,Overlap} {Area\,of\,Union}
\end{equation} 
The IoU values from tested data are used to determine the TPs and FPs. We compared YOLOs predicted bounding box the actual ground truth bounding box to obtain IoU. If the IoU is greater than 0.5, a prediction is categorized as a TP, and if it is less than 0.5, as a FP. Figure 10 illustrates the False Negative (FN) for images where the model missed an identified Type III object. 
\begin{figure}[ht]
    \centering
    \includegraphics[width=\columnwidth,height=7cm]{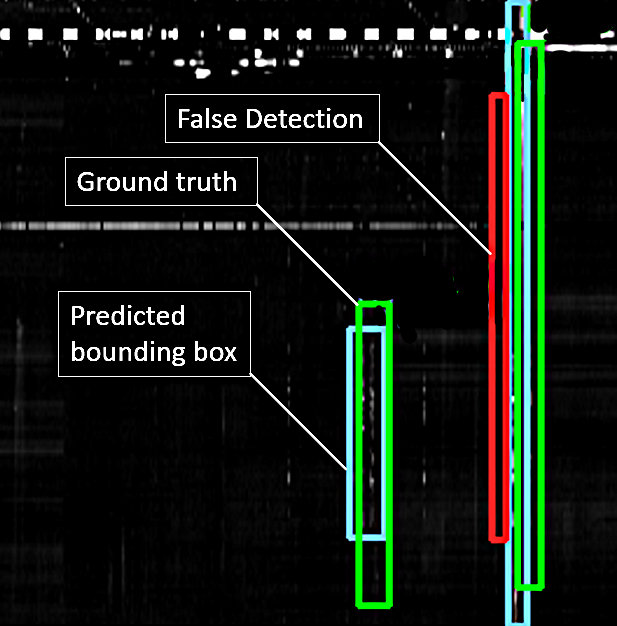}
    \caption{A visual representation of IOU thresholding. The green bounding box indicates the ground-truth or actual Type III, blue bounding box indicates a correctly predicted bounding box by YOLO (TP) and red bounding box indicates a false detection (FN) or IOU$<$0.5.}
    \label{fig:IOU_exp}
\end{figure}

The confidence threshold is an important consideration when assessing the model's performance. The confidence threshold expresses the model's level of certainty in predicting a Type III SRB. Figure 11 shows that the lower the confidence, the more detections made on a test image, but also the more false detections made. 
\begin{figure}[ht]
    \includegraphics[width=\columnwidth]{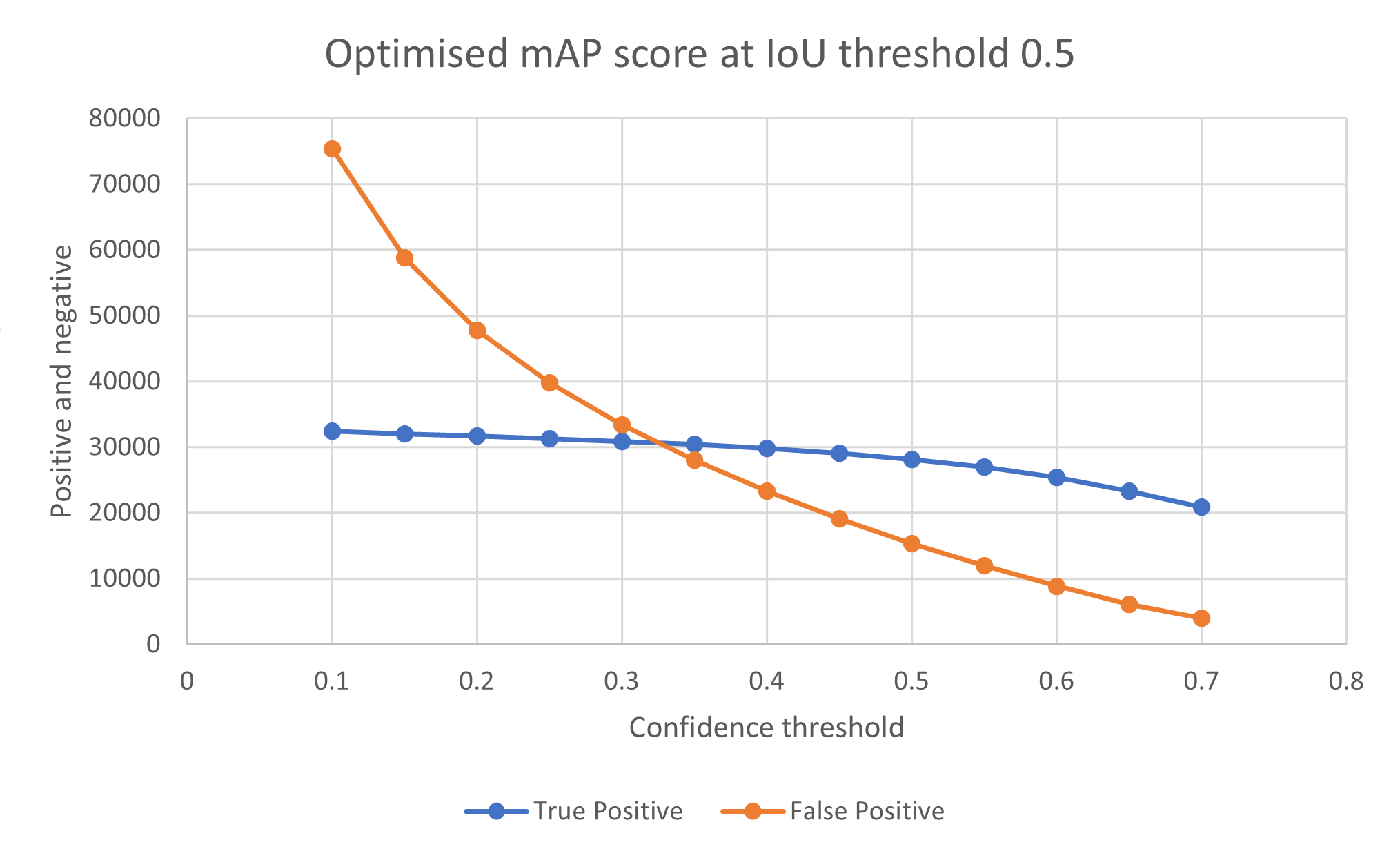}
          
\caption{A visual representation of Table 1. As the confidence threshold decreases, the TP and FP values increase. Here, we can see where YOLO performs at its optimised confidence threshold. The key is to find the balance between both true positive and true negative values. For our model, we have evaluated using IOU threshold at 0.5 (b) as it tests the model's robustness and localised accuracy at detecting Type III SRBs.
\label{fig:mAP}}
\end{figure}
The model was found to be optimal with the confidence threshold set to 0.35, as there is a balance between the TP and FP rate in terms of the models predictions (see Table 1 and Figure 11). Figure 12 shows the bounding box predictions when model configuration is set to threshold 0.35, the resulting mAP for detecting Type III solar radio bursts is 77.71\%. We plot the localised YOLO detections into a dynamic spectra observation made on the 10th of September 2017, see Figure 13.
\begin{figure*}[ht]
\centering
       \begin{subfigure}[b]{0.43\textwidth}
         \centering
         \includegraphics[width=1\textwidth]{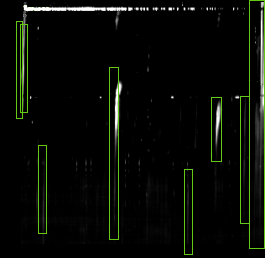}
         \caption{}
     \end{subfigure}
     \hfill
     \begin{subfigure}[b]{0.43\textwidth}
         \centering
         \includegraphics[width=1\textwidth]{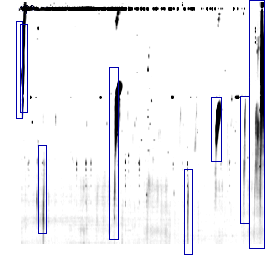}
        \caption{}
          \end{subfigure}
\caption{YOLO making localised detections on a 10-minute segment at the optimised confidence threshold of 0.35 on the testset (a). When the image is colour inverted (b), we can see the faint Type IIIs YOLO is picking up. Notice how YOLO picks up most Type IIIs in the image and ignores most RFI.
\label{fig:testset}}
\end{figure*}
\begin{figure*}[ht]
    \centering
    \includegraphics[width=1\linewidth]{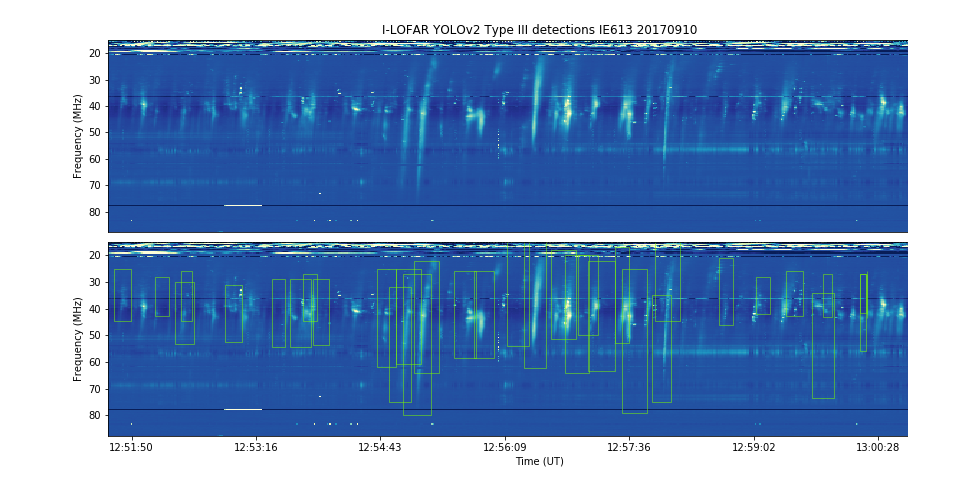}
    \caption{YOLOv2 applied to an I-LOFAR observation made on the 10th of September 2017. The models' detections capture the Type IIIs' frequency range and length in time. The model predicts the most intense Type IIIs correctly and ignores somewhat low intensity as they are quite difficult to distinguish between Type III and RFI even to the human eye.}
    \label{fig:yolovobs}
\end{figure*}

\begin{table*}[ht]
\caption{mAP scores associated with different confidence thresholds set in YOLO at different IoU thresholds. Notice how when the confidence threshold decreases the mAP increases but so too does the true negative and positive rate. The challenge is to find a balance between metrics for optimized performance in terms of accuracy.}              
\label{table:1}      
\centering                                      
\begin{tabular}{ |M{3.5cm}||M{1.6cm}|M{1.8cm}|M{1.8cm}| M{1.8cm}|M{1.6cm}|}          
\hline                        
IoU Threshold &Confidence Threshold & Recall& True Positive & False Positive &mAP \\  
\hline                                   
{     }&0.7&84.03\%& 20838 & 3959&    57.56\%\\
{     }&0.65&79.30\%& 23347 & 6093&    63.57\%\\
{     }&0.6&74.14\%& 25398 & 8858&    68.19\%\\
{     }&0.55&69.34\%& 26942 & 11911&    71.44\%\\
{     }&0.5&64.74\%& 28140 & 15321&    73.79\%\\
{     }&0.45&60.36\%& 29094 & 19099&    75.55\%\\
{     }&0.4&56.18\%& 29828 & 23261&    76.80\%\\
{IoU Threshold @ 0.5}&0.35&52.03\%& 30404 & 28026&    77.71\%\\
{     }&0.33&50.42\%& 30592 & 30073&    78.00\%\\
{     }&0.3&48.02\%& 30885 & 33422&    78.42\%\\
{     }&0.25&44.00\%& 31295 & 39827&    78.97\%\\
{     }&0.2&39.86\%& 31685 & 47801&    79.45\%\\
{     }&0.15&35.28\%& 32060 & 58811&    79.87\%\\
{     }&0.1&30.08\%& 32422 & 75336&    80.21\%\\
\hline
{     }&0.7&84.74\%& 21014 & 3783&    58.32\%\\
{     }&0.65&79.99\%& 23549 & 5891&    64.45\%\\
{     }&0.6&74.75\%& 25604 & 8648&    69.12\%\\
{     }&0.55&69.90\%& 27162 & 11691&    72.42\%\\
{     }&0.5&65.23\%& 28353 & 15108&    74.78\%\\
{     }&0.45&60.79\%& 29298 & 18895&    76.53\%\\
{     }&0.4&56.58\%& 30039 & 23050&    77.80\%\\
{IoU Threshold @ 0.1}&0.35&52.38\%& 30604 & 27822&    78.71\%\\
{     }&0.33&50.75\%& 30789 & 29876&    78.99\%\\
{     }&0.3&79.41\%& 31082 & 33225&    79.41\%\\
{     }&0.25&44.27\%& 31387 & 39635&    79.96\%\\
{     }&0.2&40.11\%& 31880 & 47606&    80.45\%\\
{     }&0.15&35.49\%& 32255 & 58616&    80.87\%\\
{     }&0.1&30.27\%& 32625 & 75133&    81.22\%\\
\hline                                             
\end{tabular}
\end{table*}

\section{Discussion}
In our previous research, we had two issues with the detection quality and robustness of the model. The first issue was detection quality, the predicted bounding boxes were very static in the y-axis or height variable. The previous parametric modelling method also lacked the ability of producing simulated Type III SRB data comparable to real observed data. With the introduction of GANs for generating Type III examples, we could produce simulated SRB data with variation in the Y-axis or height variable. GANs also provided realistic Type III examples almost identical to real observed data and also free of any interference such as RFI. The second issue robustness, where the model couldn't handle high volumes of data, having problems with SRB groupings or storms. Using the GANs simulated data, we could produce all sorts of Type III variations including groupings and storms but also other classes of Type IIIs such as inverted-U bursts and Type N bursts. Employing this new diverse training set, we could train YOLO to detect and classify the actual shape of a Type III. With this configuration of congruent deep learning models, we can accurately detect such phenomena.

In this research, we introduced the computer vision metric mAP, which calculates how well the model can accurately detect locally a Type III. One concern is the approach we took to calculate this metric. We approached the problem as a computer vision problem and evaluated it using computer vision standards applied in COCO dataset competitions, in which the IOU threshold is set to 0.5. This has many advantages in terms of accurate model detection and model robustness when tested against both busy and quiet solar activity observations. However, we could be missing out on some valuable detection data where a visually positive detection has been made (the bounding box is slightly overlapping a Type III) and the detection has been classified as a false detection due to the IoU threshold value of $<$0.5 not being met. One could argue that we have been overly harsh on the model's detections but we have done this with the view of having a more robust and accurate model for detecting Type III SRBs.

A key feature of our configuration of YOLO is the potential to detect Type III SRBs in real time. Our tests were conducted on small data streams, however, with the recent development of REALTA, a computing backend at I-LOFAR, the potential exists for the recording and processing of data in near real-time. The combination of YOLO, as a real-time software pipeline, and REALTA's hardware capabilities, could prove significant for near real-time space weather monitoring. LOFAR for Space Weather (LOFAR4SW) is a planned LOFAR improvement that would allow for frequent space weather monitoring, which will enable near-real-time monitoring of space weather phenomena such as solar flares and coronal mass ejections \citep{Carley}. This will benefit not only space weather researchers but the radio astronomy community. It will increase our understanding of how space weather affects radio wave propagation in the inner heliosphere and ionosphere disturbances, as well as the impact this has on observing astronomical sources. A backend such as REALTA will be required to capture the data streams from a LOFAR4SW-equipped international station in local mode and analyze the raw data so that it may be used by space weather researchers and forecasters. With YOLO's proven ability to detect Type IIIs in real-time, and REALTAs ability to record and process data in near real time. The possibility of near real-time Type III SRB analysis is promising.

\section{Conclusion}
We applied a combination of deep learning models to the problem of SRB generation, detection, and classification, with the focus on real-time detection. We trained a GAN to produce realistic Type III SRBs, similar to those observed in real observations from I-LOFAR. We then labeled and combined this generated data with real observed data to produce a training set on which YOLOv2 could be evaluated. This particular configuration of YOLOv2 can achieve a mAP accuracy of 77.71\% on a real data observation consisting of over 35,000 Type III solar radio burst examples while also achieving real-time frame rates (maximum 90 fps). The combination of YOLO, as a real-time software pipeline, and REALTA's hardware capabilities, could prove significant for near real-time space weather monitoring.

We intend to develop this software pipeline further by increasing the size of the dataset using GANs but also adding variety, extending YOLO's capability to detect other SRBs such as Type IIs. We have shown that with congruent deep learning model techniques, we can create a robust method of detecting Type III SRBs. Thus, illustrating that accurate real-time detection and classification of Type III SRBs is readily attainable.

\begin{acknowledgements}
    LOFAR is one of the largest astrophysics projects in Europe, consisting of 12 international stations spread across Germany, Poland, France, UK, Sweden and Ireland, with additional stations and a central hub in The Netherlands, operated by the Netherlands Institute for Radio Astronomy (ASTRON). I-LOFAR was the Irish addition to this network and was constructed by members from Trinity College Dublin (TCD), University College Dublin (UCD), Armagh Observatory, Dublin City University (DCU), University College Cork (UCC) and National University of Ireland Galway (NUIG) with funding from Science Foundation Ireland (SFI), Department of Business, Enterprise and Innovation, Open Eir and Offaly County Council. J.Scully acknowledges support from SFI and the Technological University of the Shannon (TUS).
\end{acknowledgements}


\bibliography{sample631}{}

\begin{thebibliography}{33}
\expandafter\ifx\csname natexlab\endcsname\relax\def\natexlab#1{#1}\fi

\bibitem[{Borji(2019)}]{Borji_2019}
Borji, A. 2019, Computer Vision and Image Understanding, 179, 41–65

\bibitem[{Carley {et~al.}(2020{\natexlab{a}})Carley, Baldovin, Benthem, Bisi,
  Fallows, Gallagher, Olberg, Rothkaehl, Vermeulen, Vilmer, \& et~al.}]{Carley}
Carley, E.~P., Baldovin, C., Benthem, P., {et~al.} 2020{\natexlab{a}}, Journal
  of Space Weather and Space Climate, 10

\bibitem[{Carley {et~al.}(2020{\natexlab{b}})Carley, Gallagher, Mccauley, \&
  Murphy}]{Carl}
Carley, E.~P., Gallagher, P., Mccauley, J., \& Murphy, P. 2020{\natexlab{b}},
  EGU General Assembly Conference Abstracts, 5109

\bibitem[{Connor \& van Leeuwen(2018)}]{Connor}
Connor, L. \& van Leeuwen, J. 2018, The Astronomical Journal, 156, 256

\bibitem[{Evgeniou \& Pontil(2001)}]{SVM}
Evgeniou, T. \& Pontil, M. 2001, Lecture Notes in Computer Science (including
  subseries Lecture Notes in Artificial Intelligence and Lecture Notes in
  Bioinformatics), 2049 LNAI, 249–257

\bibitem[{Girshick(2015)}]{Fast_RCNN}
Girshick, R. 2015, Proceedings of the IEEE International Conference on Computer
  Vision, 2015 Inter, 1440–1448

\bibitem[{Girshick {et~al.}(2016)Girshick, Donahue, Darrell, \& Malik}]{R-CNN}
Girshick, R., Donahue, J., Darrell, T., \& Malik, J. 2016, IEEE Transactions on
  Pattern Analysis and Machine Intelligence, 38, 142–158

\bibitem[{Goodfellow(2016)}]{Goodfellow_2016}
Goodfellow, I. 2016

\bibitem[{Goodfellow {et~al.}(2020)Goodfellow, Pouget-Abadie, Mirza, Xu,
  Warde-Farley, Ozair, Courville, \& Bengio}]{Gans}
Goodfellow, I., Pouget-Abadie, J., Mirza, M., {et~al.} 2020, Communications of
  the ACM, 63, 139–144

\bibitem[{He {et~al.}(2016)He, Zhang, Ren, \& Sun}]{He_Zhang}
He, K., Zhang, X., Ren, S., \& Sun, J. 2016, Proceedings of the IEEE Computer
  Society Conference on Computer Vision and Pattern Recognition, 2016-Decem,
  770–778

\bibitem[{Hou {et~al.}(2020)Hou, Zhang, Feng, Du, Gao, Zhao, \&
  Miao}]{Faster_RCNN_SRB}
Hou, Y.~C., Zhang, Q.~M., Feng, S.~W., {et~al.} 2020, Solar Physics, 295

\bibitem[{Kalkan {et~al.}(2018)Kalkan, Okur, \& Altunışık}]{parametric}
Kalkan, E., Okur, F., \& Altunışık, A. 2018, Journal of Construction
  Engineering, Management and Innovation, 1, 139–146

\bibitem[{Lin(2011)}]{Lin_2011}
Lin, R.~P. 2011, Space Science Reviews, 159, 421

\bibitem[{Liu {et~al.}(2016)Liu, Anguelov, Erhan, Szegedy, Reed, Fu, \&
  Berg}]{SSD}
Liu, W., Anguelov, D., Erhan, D., {et~al.} 2016, Lecture Notes in Computer
  Science (including subseries Lecture Notes in Artificial Intelligence and
  Lecture Notes in Bioinformatics), 9905 LNCS, 21–37

\bibitem[{Lobzin {et~al.}(2014)Lobzin, Cairns, \& Zaslavsky}]{Lobzin}
Lobzin, V.~V., Cairns, I.~H., \& Zaslavsky, A. 2014, Journal of Geophysical
  Research A: Space Physics, 119, 742–750

\bibitem[{Louppe(2014)}]{RF}
Louppe, G. 2014, Machine Learning

\bibitem[{Lu {et~al.}(2004)Lu, Wang, Dizaji, Ding, \& Ponsford}]{CRAF}
Lu, X., Wang, J., Dizaji, R., Ding, Z., \& Ponsford, A.~M. 2004, Canadian
  Conference on Electrical and Computer Engineering, 4, 2081–2084

\bibitem[{Ma {et~al.}(2017)Ma, Chen, Xu, \& Yan}]{Multimodal}
Ma, L., Chen, Z., Xu, L., \& Yan, Y. 2017, Pattern Recognition, 61, 573–582

\bibitem[{Murphy {et~al.}(2021)Murphy, Callanan, McCauley, McKenna,
  Fionnagáin, Louis, Redman, Cañizares, Carley, Maloney, \& et~al.}]{Murphy}
Murphy, P.~C., Callanan, P., McCauley, J., {et~al.} 2021, Astronomy and
  Astrophysics

\bibitem[{Pick~M(2009)}]{Pick}
Pick~M, V.~N. 2009, The Astronomy and Astrophysics Review

\bibitem[{Rafegas \& Vanrell(2017)}]{colour}
Rafegas, I. \& Vanrell, M. 2017, Proceedings - 2017 IEEE International
  Conference on Computer Vision Workshops, ICCVW 2017, 2018-Janua, 2697–2705

\bibitem[{Redmon {et~al.}(2016)Redmon, Divvala, Girshick, \& Farhadi}]{YOLO}
Redmon, J., Divvala, S., Girshick, R., \& Farhadi, A. 2016, Proceedings of the
  IEEE Computer Society Conference on Computer Vision and Pattern Recognition,
  2016-Decem, 779–788

\bibitem[{Redmon \& Farhadi(2017)}]{YOLO9000}
Redmon, J. \& Farhadi, A. 2017, Proceedings - 30th IEEE Conference on Computer
  Vision and Pattern Recognition, CVPR 2017, 2017-Janua, 6517–6525

\bibitem[{Reid \& Ratcliffe(2014)}]{typeoccurence}
Reid, H. A.~S. \& Ratcliffe, H. 2014, 14, 773–804

\bibitem[{Ren {et~al.}(2017)Ren, He, Girshick, \& Sun}]{Faster_RCNN}
Ren, S., He, K., Girshick, R., \& Sun, J. 2017, IEEE Transactions on Pattern
  Analysis and Machine Intelligence, 39, 1137–1149

\bibitem[{Scully {et~al.}(2021)Scully, Flynn, Carley, Gallagher, \&
  Daly}]{Scully_Flynn_Carley_Gallagher_Daly_2021}
Scully, J., Flynn, R., Carley, E., Gallagher, P., \& Daly, M. 2021, in Irish
  Signal and Systems Conference 2021, 1–6

\bibitem[{Trieu(2018)}]{darkflow}
Trieu, T.~H. 2018, GitHub Repository. Available online: https://github.
  com/thtrieu/darkflow (accessed on 14 February 2019)

\bibitem[{Van~Haarlem {et~al.}(2013)Van~Haarlem, Wise, Gunst, Heald, McKean,
  Hessels, De~Bruyn, Nijboer, Swinbank, Fallows, \& et~al.}]{Van_Haarlem}
Van~Haarlem, M.~P., Wise, M.~W., Gunst, A.~W., {et~al.} 2013, Astronomy and
  Astrophysics, 556

\bibitem[{Wu(2017)}]{CNN}
Wu, J. 2017, Introduction to Convolutional Neural Networks, 1–31

\bibitem[{Ying(2019)}]{overfit}
Ying, X. 2019, Journal of Physics: Conference Series, 1168

\bibitem[{Zhang {et~al.}(2021)Zhang, Yan, Han, He, Li, Wu, \& Chen}]{DCGAN_SRB}
Zhang, W., Yan, F., Han, F., {et~al.} 2021, Frontiers in Physics, 9, 1–8

\bibitem[{Zhang {et~al.}(2018)Zhang, Gajjar, Foster, Siemion, Cordes, Law, \&
  Wang}]{Seti}
Zhang, Y.~G., Gajjar, V., Foster, G., {et~al.} 2018, arXiv

\bibitem[{Zhuang {et~al.}(2021)Zhuang, Qi, Duan, Xi, Zhu, Zhu, Xiong, \&
  He}]{transfer}
Zhuang, F., Qi, Z., Duan, K., {et~al.} 2021, Proceedings of the IEEE, 109,
  43–76

\end{thebibliography}
\bibliographystyle{aa}

\end{document}